\documentclass[prl,preprint,showpacs,aps]{revtex4}     
\usepackage{amsmath}
\def\beq{\begin{equation}}
\def\eeq{\end{equation}}
\def\bea{\begin{eqnarray}}
\def\eea{\end{eqnarray}}
\def\nnu{\nonumber}
\def\tst{\textstyle}

\def\eno#1{Eq.~(\ref{#1})}
\def\Eno#1{Equation (\ref{#1})}
\def\etwo#1#2{Eqs.~(\ref{#1}) and (\ref{#2})}
\def\Etwo#1#2{Equations (\ref{#1}) and (\ref{#2})}

\def\gtwid{\mathrel{\raise.3ex\hbox{$>$\kern75em\lower1ex\hbox{$\sim$}}}}
\def\ltwid{\mathrel{\raise.3ex\hbox{$<$\kern-.75em\lower1ex\hbox{$\sim$}}}}

\def\al{\alpha}
\def\be{\beta}
\def\gam{\gamma}
\def\dta{\delta}
\def\eps{\epsilon}
\def\ze{\zeta}
\def\tta{\theta}

\def\lam{\lambda}
\def\sig{\sigma}

\def\vph{\varphi}

\def\Dta{\Delta}

\def\bsig{{\boldsymbol{\sigma}}}
\def\btau{{\boldsymbol{\tau}}}

\def\apx{\approx}
\def\ptl{\partial}
\def\hf{\frac{1}{2}}
\def\tshf{{\tst\hf}}

\def\qua{\frac{1}{4}}
\def\tsqua{{\tst\qua}}

\def\dint{\int\!\!\!\int}

\def\grad{\nabla}

\def\part#1#2{\frac{\ptl#1}{\ptl#2}}

\def\ket#1{|#1\rangle}

\def\bra#1{\langle#1|}

\def\olap#1#2{\langle#1|#2\rangle}
\def\avg#1{\langle#1\rangle}
\def\mel#1#2#3{\langle#1|#2|#3\rangle}

\def\bavg#1{\bigl\langle#1\bigr\rangle}
\def\Bavg#1{\left\langle#1\right\rangle}
\def\tr{{\rm tr}\,}

\def\ba{{\bf a}}

\def\bj{{\bf j}}

\def\br{{\bf r}}
\def\bs{{\bf s}}
\def\bt{{\bf t}}

\def\bJ{{\bf J}}

\def\xhat{{\bf{\hat x}}}
\def\yhat{{\bf{\hat y}}}
\def\zhat{{\bf{\hat z}}}
\def\nhat{{\bf{\hat n}}}

\def\jtil{\tilde{\jmath}}
\def\Fe8{Fe$_8$}
\def\Mn12{Mn$_{12}$}

\def\am{{\cal L}}
\def\rmap{{\cal R}}
\def\ylm{Y_{\ell m}}
\def\Ylm{{{\cal Y}}_{\ell m}}

\def\qlm{\Phi^{Q}_{\ell m}}
\def\plm{\Phi^{P}_{\ell m}}
\def\wlm{\Phi^{W}_{\ell m}}

\begin{document}


\title{The Weyl-Wigner-Moyal Formalism for Spin}
\author{Feifei Li}
\author{Carol Braun$^{\dagger}$}
\author{Anupam Garg}
\email[e-mail address: ]{agarg@northwestern.edu}
\affiliation{Department of Physics and Astronomy, Northwestern University,
Evanston, Illinois 60208}

\date{\today}

\begin{abstract}
The Weyl-Wigner-Moyal formalism is developed for spin by means of a correspondence between spherical harmonics and spherical
harmonic tensor operators. The exact asymptotic relation between the P, Q, and Weyl symbols is found, and the analogue of the
Moyal expansion is developed for the Weyl symbol of the product of two operators in terms of the symbols for the individual
operators. It is shown that in the classical limit, the 
Weyl symbol for a commutator equals $i$ times the Poisson bracket of the corresponding Weyl symbols.
\end{abstract}

\maketitle

For a particle with linear momentum $p$ and coordinate $q$, the Weyl-Wigner-Moyal
formalism~\cite{weyl,wig,moy} provides one of the most powerful ways of understanding the relation between classical
and quantum mechanics, as it leads to an explicit
mapping between quantum mechanical operators and classical dynamical variables defined as functions on phase space.
When the operator is the density matrix, the phase space function is known as the Wigner function. Because this function
fails to be nonnegative in all circumstances, it cannot be taken to be a true probability distribution. So while it does nothing
to solve the knotty problem of interpreting quantum mechanics, this formalism is the closest one can get to turning quantum
mechanics into a purely classical statistical theory, and it also finds practical application, especially in quantum optics.

It is desirable to
have a similar formalism for spin degrees of freedom both for intrinsic reasons, and for developing the semiclassical limit.
In their low energy states, many atoms, molecules, and molecular ions behave as particles with a fixed magnitude of spin
that is often large, in which case a semiclassical approach is natural. We have found, for example, that the Weyl
representation is advantageous in constructing the spin-coherent-state path integral~\cite{tbp}.
The problem of developing such a formalism has been approached by several
workers~\cite{Strato56,Bayen78,Girish81,vgb89,chumakov00} with varying emphases and methods. The problem is solved in a formal
sense and its abstract mathematical aspects are quite highly developed. At the same time, unlike the case of systems with
$p$ and $q$ variables, there are few closed-form results of practical value. One of the loveliest results in the semiclassics of
$p$ and $q$ systems is that the leading $\hbar \to 0$  behavior of the classical object corresponding to the commutator
(known as the Moyal bracket) is the same as the Poisson bracket.
We show for the first time in this paper that the same statement holds for spin as the spin magnitude $j \to \infty$, even though
the actual form of the Poisson bracket is rather different. While one might have guessed this fact heuristically, that is not the
same as proving it, and so far as rigor in one's conception of physics is valuable, so is a proof. One can expect terms of higher
order in $1/j$ to aid in developing quantum corrections as in Ref.~\cite{wig}. 

We also find the {\it exact\/} asymptotic relation between the P, Q, and Weyl symbols of spin operators as
$j \to \infty$. This is done by finding a novel projection of the sphere onto the plane. While the same projection is also
useful in finding the Moyal product, explicit derivation of the asymptotic series is very tedious, and only the leading term
mentioned above is easily written.

The objects of interest to us are operators for a particle of spin $j$, which are functions of $\bJ$, the vector spin operator,
with components $J_x$, $J_y$, and $J_z$ that have commutators
$[J_{\al}, J_{\be}] = i \eps_{\al\be\gam} J_{\gam}$ (setting $\hbar$ to 1). Also,
$
\bJ\cdot\bJ = J_x^2 + J_y^2 + J_z^2 = j(j+1)
$.
For any such operator $A(\bJ)$, the P and Q symbols are defined by \cite{nonunique}
\bea
\Phi^Q_A(\nhat) &=& \mel{\nhat}{A(\bJ)}{\nhat}, \label{def_qsym} \\
\noalign{\vskip4pt}
A(\bJ) &=& \frac{2j+1}{4\pi} \int d\nhat\, \ket{\nhat} \Phi^P_A(\nhat) \bra{\nhat}. \label{def_psym}
\eea
(The Weyl symbol, $\Phi^W_A$, will be defined below.)
Here, $\ket{\nhat}$ is the spin state that is maximally aligned along the direction $\nhat$. That is,
$
\bJ\cdot\nhat\, \ket{\nhat} = j \ket{\nhat}
$,
and $\olap{\nhat}{\nhat} = 1$. Thus, if $\ket{j,m}$ denotes the simultaneous eigenstate of $\bJ\cdot\bJ$ and $J_z$,
$
\ket{j, j} \equiv \ket{\zhat}
$.
In \eno{def_psym}, the integral is over all $\nhat$.
It also pays to employ stereographic coordinates.
If $\tta$ and $\vph$ denote the spherical polar coordinates of $\nhat$, and we define
$
z = \tan \tshf\tta e^{i\vph}
$,
and
\beq
\ket{z} = e^{zJ_-} \ket{j,j}
        = \sum_{m = -j}^j {\binom{2j}{j-m}}^{1/2} z^{j-m}\, \ket{j, m}, \label{def_z_state}
\eeq
then it is a standard result that
$
\ket{z} = (1 + |z|^2)^j\, \ket{\nhat}
$.

A physically meaningful phase space representation must be linear, real for Hermitean
operators, equal to 1 for the identity operator, and covariant under rotations. That is, if the symbol for
$A(\bJ)$ is $\Phi_A(\nhat)$, and if a rotation takes $\bJ$ into $\bJ'$ and $\nhat$ into $\nhat'$, then the
symbol for $A(\bJ')$ must be $\Phi_A(\nhat')$. These properties are obeyed by $\Phi^Q_A$ and $\Phi^P_A$.
The Weyl symbol must also obey
the key extra demand of {\it traciality\/}~\cite{Strato56}. That is, if $A$ and $B$ are two operators, we must have
\beq
\frac{1}{2j+1} \tr(AB) = \frac{1}{4\pi} \int d\nhat\, \Phi^W_A(\nhat) \Phi^W_B(\nhat). \label{tr_weyl}
\eeq
The left and right sides of this equation may be seen as quantum and classical averages, respectively.
We denote these as $\avg{\ \ }_{\rm qm}$ and $\avg{\ \ }_{\nhat}$. This condition determines $\wlm(\nhat)$ unambiguously.
Further, if $B$, say, is taken as the density matrix, then $(2j+1) \Phi^W_B(\nhat)/4\pi$ is the Wigner function for the system. 

We first consider the P, Q, and W symbols (denoted $\Phi_{\ell m}^P$ etc.) of the spherical harmonic tensor operators
$\Ylm(\bJ)$. We find that
for a particle of spin $j$,
\beq
\Phi_{\ell m}^{P,\, Q,\, W}(\nhat) = a^{P,\, Q,\, W}_{j \ell} \ylm(\nhat), \label{def_a_coeffs}
\eeq
where $\ylm(\nhat)$ is a spherical harmonic, and
\bea
a^P_{j \ell} &=& \bigl(j + 1\bigr) \bigl(j +{\tst{\frac{3}{2}}})
                                        \bigl(j + 2\bigr) \cdots \bigl(j + \tshf\ell + \tshf \bigr), \label{answer_plm}\\
a^Q_{j \ell} &=& j \bigl(j - \tshf\bigr) \bigl(j - 1\bigr) \cdots \bigl(j - \tshf\ell + \tshf \bigr), \label{answer_qlm}\\
a^W_{j \ell} &=& \bigl(a^P_{j \ell} a^Q_{j \ell} \bigr)^{1/2}
               =  \prod_{k = 1}^{\ell} \bigl((j +\tshf)^2 -\tsqua k^2 \bigr)^{1/2}.
   \label{answer_wlm}
\eea
Since any operator can be uniquely expressed as a linear combination of the $\Ylm(\bJ)$'s, these results
constitute an algorithmic solution to the problem of finding the P, Q, and Weyl representations for an arbitrary operator.

To prove Eqs.~(\ref{def_a_coeffs})--(\ref{answer_wlm}), we 
define $\ylm$ and $\Ylm(\bJ)$ via the Herglotz generating function~\cite{ag_embook}:
\bea
\!\!\!\!e^{\zeta \ba\cdot\br}
   &=& \sum_{\ell, m} \sqrt{\frac{4\pi}{2\ell + 1}}\, 
      \frac{r^{\ell} \zeta^{\ell} \lam^m}{\sqrt{(\ell + m)! (\ell -m)!}} \ylm(\nhat), \label{herglotz} \\
\!\!\!\!e^{\zeta \ba\cdot\bJ}
   &=& \sum_{\ell, m} \sqrt{\frac{4\pi}{2\ell + 1}}\, 
      \frac{\zeta^{\ell} \lam^m}{\sqrt{(\ell + m)! (\ell -m)!}} \Ylm(\bJ). \label{herg2}
\eea
Here,
\beq
\ba = \zhat - \frac{\lam}{2} (\xhat + i \yhat) + \frac{1}{2\lam} (\xhat - i \yhat),
\eeq
with $\lam$ and $\zeta$ being real. The essential property of $\ba$ is that
$
\ba\cdot\ba = 0
$.
Further, $r = |\br|$, and $\nhat = \br/r$. \Eno{herglotz} implies the normalization $\int d^2\nhat\, |\ylm|^2 = 1$.
\Etwo{herglotz}{herg2} show that various relationships among the $\ylm$'s have direct analogs for the $\Ylm$'s. For example,
just as
$
\ylm^*(\nhat) = (-1)^m Y_{\ell,-m}(\nhat)
$,
so,
$
\Ylm^{\dagger}(\bJ) = (-1)^m {{\cal Y}}_{\ell,-m}(\bJ)
$,
Further, it is apparent that under rotations $\ylm(\nhat)$ and $\Ylm(\bJ)$ transform identically.

We now note that $\Ylm(\bJ)$ is the operator analogue of the {\it solid\/} harmonic $r^{\ell} \ylm(\nhat)$ with an
extra factor $r^{\ell}$ over the {\it surface\/} harmonic $\ylm$. Hence, we expect all three symbols for
$\Ylm(\bJ)$ to asymptote to $j^{\ell} \ylm(\nhat)$ as $j \to \infty$.
Second, for spin $j$, any operator is equivalent to a matrix of order $2j+1$. A Hermitean matrix of this order
has $(2j+1)^2$ independent parameters, so there can be at most $(2j+1)^2$ independent $\Ylm$'s.
This number is exhausted by taking $\ell$ up to $2j$, and the rotational properties then guarantee that
$\Ylm(\bJ) = 0$ if $\ell > 2j$.

We now find $\qlm$. Taking the expectation value of \eno{herglotz} in the state $\ket{\nhat}$ yields
\beq
\mel{\nhat}{e^{\zeta \ba\cdot\bJ}}{\nhat}
   = \sum_{\ell, m} \sqrt{\frac{4\pi}{2\ell + 1}}\, 
      \frac{\zeta^{\ell} \lam^m}{\sqrt{(\ell + m)! (\ell -m)!}} \qlm(\nhat).  \label{herg_avg}
\eeq
Since $\ket{\nhat} = (1 +|z|^2)^{-j} \ket{z}$,
\bea
\!\!\!\!\!\!
\mel{\nhat}{e^{\zeta \ba\cdot\bJ}}{\nhat}
   &\!\!=\!\!& (1 + |z|^2)^{-2j} \mel{z}{e^{\zeta \ba\cdot\bJ}}{z} \nnu\\
   &\!\!=\!\!& (1 + |z|^2)^{-2j} \mel{j,j}{e^{z^* J_+} e^{\zeta \ba\cdot\bJ} e^{zJ_-}}{j,j}. \label{avg_exp_J}
\eea
The Lie algebra of the $J_{\al}$'s allows us to write
$
e^{z^* J_+} e^{\zeta \ba\cdot\bJ} e^{zJ_-} = e^{u_-J_-} e^{\be J_z} e^{u_+ J_+}
$,
where $u_{\pm}$ and $\be$ are functions of $\zeta$, $z^*$, $z$, and $\ba$. Hence,
$
\mel{z}{e^{\zeta \ba\cdot\bJ}}{z} = e^{\be j}
$,
and
$
\mel{\nhat}{e^{\zeta \ba\cdot\bJ}}{\nhat} = e^{\gam j}
$,
where
$
e^{\gam} = e^{\be}/(1+|z|^2)^2
$.
To find $\gam$, we exploit the faithfulness of the $j=1/2$ representation of SU(2). Then,
$
(\ba\cdot\bJ)^2 = a_i a_j (\dta_{ij} + 2i \eps_{ijk} J_k)/4 = 0
$,
so,
$
e^{\zeta \ba\cdot\bJ} = 1 + \zeta \ba\cdot\bJ
$.
Now, $\mel{\nhat}{\bJ}{\nhat} = j\nhat$, so for $j=1/2$,
$
\mel{\nhat}{e^{\zeta \ba\cdot\bJ}}{\nhat} = 1 + \tshf\ze \ba\cdot\nhat = e^{\gam/2}
$.
Hence, for general $j$,
\beq
\mel{\nhat}{e^{\ze\ba\cdot\bJ}}{\nhat}
  = \bigl(1 + \tshf \ze\ba\cdot\nhat\bigr)^{2j} \label{q_rep_exp}
  = \sum_{\ell = 0}^{2j} {2j\choose\ell} \frac{\ze^{\ell}}{2^{\ell}} (\ba\cdot\nhat)^{\ell}.
\eeq
If we now 
use the $\zeta^{\ell}$ term in \eno{herglotz} for $(\ba\cdot\nhat)^{\ell}$, and compare with \eno{herg_avg}, we obtain
\beq
\qlm(\nhat) = \frac{1}{2^{\ell}} \frac{ (2j)!}{ (2j - \ell)!} \ylm(\nhat)
                  \quad  (\ell \le 2j), \label{result_qlm}
\eeq
which is the same as \eno{answer_qlm}.
As expected, $\qlm(\nhat)$ is proportional to $\ylm(\nhat)$, and it vanishes if $\ell > 2j$.

Knowing $\qlm(\nhat)$, $\plm$ is simple to find. Taking the expectation value of \eno{def_psym} in the state
$\ket{\nhat}$, and noting that
$
|\olap{\nhat}{\nhat'}|^2 = [(1+\nhat\cdot\nhat')/2]^{2j}
$,
we obtain
\beq
\Phi^Q_A(\nhat) = \frac{2j+1}{4\pi} \int d\nhat'\, \Bigl(\frac{1 + \nhat\cdot\nhat'}{2} \Bigr)^{2j} \Phi^P_A(\nhat').
    \label{prep_to_qrep}
\eeq
It is straightforward to show that
\beq
\Bigl(\frac{1 + \nhat\cdot\nhat'}{2} \Bigr)^{2j} = \sum_{\ell=0}^{2j} K_{j \ell} P_{\ell}(\nhat\cdot\nhat'),
      \label{kernel_expand}
\eeq
where
\beq
K_{j\ell} = (2\ell + 1) \frac{ \bigl[(2j)!\bigr]^2}{(2j-\ell)!\, (2j + \ell +1)!}, \label{kjl}
\eeq
and $P_{\ell}$ is the Legendre polynomial of order $\ell$. We take $A = \Ylm(\bJ)$, and feed $\plm$ and $\qlm$ from
Eqs.~(\ref{def_a_coeffs})--(\ref{answer_qlm})
along with \eno{kernel_expand} into \eno{prep_to_qrep}, and invoke the addition theorem and orthonormality
of the $\ylm$'s to obtain
\beq
a^Q_{j \ell} = \frac{2j+1}{2\ell+1} K_{j\ell} \, a^P_{j\ell}.
\eeq
Using \eno{result_qlm} for $a^Q_{j \ell}$ yields \eno{answer_plm} for $a^P_{j \ell}$.

We now find $\Phi^W_{\ell m}(\nhat)$. We evaluate the trace in \eno{tr_weyl} by writing \eno{def_psym} for $A$, and multiply
by $B$ to obtain
\beq
\Bavg{ \Phi^W_A(\nhat) \Phi^W_B(\nhat)}_{\nhat}
  = \avg{AB}_{{\rm qm}}
  = \bavg{ \Phi^P_A(\nhat) \Phi^Q_B(\nhat)}_{\nhat}.
 \label{ww_pq}
\eeq
Naturally, we can switch the roles of P and Q on the right. If we take $\Ylm$ for $A$ and ${\cal Y}_{\ell' m'}$ for $B$,
we find \eno{answer_wlm} for $a^W_{j\ell}$ when $\ell = \ell'$, More generally,
\beq
\bavg{ \Ylm(\bJ) {\cal Y}_{\ell' m'}^{\dagger}(\bJ)}_{{\rm qm}}
   = \frac{1}{4\pi} \bigl(a^W_{j \ell}\bigr)^2
        \dta_{\ell\ell'}\dta_{mm'},
       \label{tr_yy}
\eeq
which is an operator orthogonality relation. In particular,
$
a^W_{j 1} = [j(j+1)]^{1/2}
$,
so that for $\bJ$ itself, the Weyl symbol is $\sqrt{j(j+1)}\, \nhat$. This is a direct deduction of exactly
what is indicated for the ``classical" vector corresponding to $\bJ$ by a large number of indirect quantum mechanical arguments.
Further, for fixed $\ell$, we find that
\beq
\wlm(\bJ) \apx \Bigl(1 + \frac{\ell}{2j} + \cdots\Bigr) j^{\ell} \ylm(\nhat), \quad (j\to\infty).
\eeq

We now study how $\Phi^P_A$ and $\Phi^Q_A$ are related as $j \to \infty$. To this end, we rewrite 
\eno{prep_to_qrep} as follows. Let $\nhat'$ have spherical polar coordinates $\tta$ and
$\vph$ taking $\nhat$ as the pole. We map this onto a vector $\bs$ in a two-dimensional plane by using
plane polar coordinates $(s,\vph_s)$ with
\beq
e^{-s^2/4} = \cos^2 \tshf\tta, \quad \vph_s = \vph.   \label{exp_map}
\eeq
Then, $d^2\nhat = e^{-s^2/4} d^2s$, and with $\jtil = j + \tshf$,
\beq
\Phi^Q_A(\nhat) = \frac{\jtil}{2\pi} \int e^{-\jtil s^2/2} \Phi^P_A \bigl(\nhat'(\nhat,\bs)\bigr)\, d^2s.
\eeq
The integral is dominated by small $s$ for large $\jtil$. Hence, we may Taylor expand $\Phi^P_A$ in powers of
$\bs$, after which the integral is easy, and gives the exact asymptotic relation,
\beq
\Phi^Q_A(\nhat) \apx e^{\grad_s^2/2\jtil} \left. \Phi^P_A \bigl(\nhat'(\nhat,\bs)\bigr) \right|_{\bs = 0}.
     \label{psym_to_qsym}
\eeq

Let us write \eno{psym_to_qsym} symbolically as $\Phi^Q_A(\nhat) = \rmap \Phi^P_A(\nhat)$. The operator $\rmap$
must be expressible entirely in terms of $\am^2$, where
$
\am = - i (\nhat \times \grad_{\nhat})
$
is the angular momentum operator (on phase space, and {\it not\/} the quantum mechanical Hilbert space), as
that is the fundamental scalar operator present. For small $s$, $\bs \apx \nhat' - \nhat$, so
$\grad_s^2 \apx -\am^2$. We can develop $\rmap$ as a series in $1/\jtil$ to find~\cite{easier}
\beq
\rmap = 1 - \frac{\am^2}{2\jtil} + \frac{\am^4}{8\jtil^2} - \frac{\am^4(\am^2 + 1)}{48\jtil^3} + \cdots.
\eeq
It follows from \eno{answer_wlm} that $\Phi^W_A = \rmap^{1/2}\Phi^P_A = \rmap^{-1/2}\Phi^Q_A$. Clearly,
$\rmap^{\pm 1/2}$ may also be found as a series in $1/\jtil$.

In the rest of this paper, we consider the Moyal product, i.e., the Weyl symbol for the product of two operators
$AB$ in terms of $\Phi^W_A$ and $\Phi^W_B$. For this, we first introduce the Stratonovich-Weyl kernel
$\Dta^j(\nhat)$~\cite{Strato56,vgb89}, which is is an operator valued function of $\nhat$ with the
properties
\beq
\Phi^W_A(\nhat) = \bavg{ A \Dta^j(\nhat)}_{{\rm qm}}, \quad
A =
\bavg{\Phi^W_A(\nhat) \Dta^j(\nhat)}_{\nhat}.
\eeq
It is straightforward to show that
\beq
\Dta^j = 4\pi \sum_{\ell, m} \ylm^*(\nhat) \Ylm(\bJ)/a^W_{j\ell}. \label{answer_Dta}
\eeq
In Ref.~\cite{vgb89} instead, $\Dta^j$ is given as a matrix in the $\ket{j,m}$ basis.
It is easily seen that $\Dta^j$ must obey the rules,
\beq
\avg{\Dta^j(\nhat)}_{\nhat} = 1, \quad
\bavg{ \Dta^j(\nhat) \Dta^j(\nhat') }_{{\rm qm}} = 4\pi I^j(\nhat,\nhat'). \label{tr_Dta_Dta}
\eeq
Here, $I^j(\nhat,\nhat')$ is the identity kernel on the space of harmonic functions of $\nhat$
of order $2j$ or less. That is,
\beq
\int d\nhat'\, I^j(\nhat, \nhat') f(\nhat') = f(\nhat)
\eeq
for any function $f(\nhat)$ whose spherical harmonic expansion contains no terms with $\ell > 2j$.
The properties (\ref{tr_Dta_Dta}) follow from \etwo{answer_Dta}{tr_yy}.

Using the $\Dta^j$ operator, one finds that~\cite{vgb89}
\bea
&&\Phi^W_{AB}(\nhat_1)
   = \bavg{M_j(\nhat_1, \nhat_2, \nhat_3) \Phi^W_A(\nhat_2) \Phi^W_B(\nhat_3) }_{\nhat_2, \nhat_3}; \\
&&M_j(\nhat_1, \nhat_2, \nhat_3)
   = \bavg{ \Dta^j(\nhat_1) \Dta^j(\nhat_2) \Dta^j(\nhat_3)}_{{\rm qm}}.
\eea
As $j \to \infty$, the trikernel $M_j$ is very sharply peaked when $\nhat_1 = \nhat_2 = \nhat_3$ and nearly zero
otherwise. Further, progressively higher moments of $M_j$ are of higher order in $1/j$, suggesting that an asymptotic
series for $\Phi^W_{AB}$ may be found by Taylor expanding $\Phi^W_A$ and $\Phi^W_B$ in the angles
$\cos^{-1}(\nhat_1\cdot\nhat_2)$ and $\cos^{-1}(\nhat_1\cdot\nhat_3)$. While this is correct, the lack of an explicit
form for $M_j$ makes it very cumbersome. A better procedure is as follows. We write
\beq
\Phi^Q_{AB}(\nhat_1) = \mel{\nhat_1}{AB}{\nhat_1}.
\eeq
Writing $A$ and $B$ in the P representation gives
\beq
\Phi^Q_{AB}(\nhat_1)
   = \frac{\jtil^2}{4\pi^2} \dint d^2\nhat_2\, d^2\nhat_3\,
        T_j(\nhat_1, \nhat_2, \nhat_3) \Phi^P_A(\nhat_2) \Phi^P_B(\nhat_3),
     \label{pp_to_q}
\eeq
where
\beq
T_j(\nhat_1, \nhat_2, \nhat_3)
  = \olap{\nhat_1}{\nhat_2} \olap{\nhat_2}{\nhat_3} \olap{\nhat_3}{\nhat_1}.
\eeq
A short calculation shows that $T_j = T_{1/2}^{2j}$, where
\beq
T_{1/2}
  = \frac{1 + \nhat_1\cdot\nhat_2 + \nhat_2\cdot\nhat_3
            + \nhat_3\cdot\nhat_1 + i\nhat_1\cdot(\nhat_2\times \nhat_3)}{4}.
\eeq
The trikernel $T_j$ is also sharply peaked at $\nhat_1 = \nhat_2 = \nhat_3$, and very small otherwise.
Thus, \eno{pp_to_q} can also be the basis of an asymptotic expansion similar to that leading to
\eno{psym_to_qsym}. To do this, we map $\nhat_2$ to $\bs$ as before using \eno{exp_map}, and also to
another two-dimensional vector $\bsig$ via Lambert's equal area projection. That is, if $\bsig$ has plane polar
coordinates $(\sig, \vph_{\sig})$, and $\nhat_2$ has spherical polar coordinates $(\tta_2, \vph_2)$ with $\nhat_1$
as the pole, then $\sig = 2\sin\tshf\tta_2$, and $\vph_{\sig} = \vph_2$. The corresponding maps for $\nhat_3$ take
us to vectors $\bt$ and $\btau$. We have
\beq
\sig^2 = s^2 - \frac{s^4}{8} + \frac{s^6}{96} + \cdots,
\eeq
and likewise for $\tau^2$ in terms of $t^2$. In terms of these vectors, we find
\beq
T_{j}\,d^2\nhat_2\, d^2\nhat_3
  = \exp \Bigl[-\tshf \jtil\, Q'(\bs,\bt) \Bigr] d^2s\, d^2t,
\eeq
with
\bea
Q'(\bs,\bt) &=& s^2 + t^2 - 4\frac{j}{\jtil} 
          \ln \Bigl(1 + \frac{\sig\tau}{4st} e^{(s^2 + t^2)/8} v(\bs, \bt) \Bigr), \\
v(\bs,\bt) &=& \bs\cdot\bt + i \nhat_1\cdot(\bs\times\bt).
\eea
Thus $T_j$ is almost a Gaussian, with widths in $s$ and $t$ of order $j^{-1/2}$. Hence, the logarithm in $Q'$ may be expanded
to yield $Q'(\bs,\bt) = Q(\bs,\bt) - R_4 - R_6 - \cdots$, where
\beq
Q(\bs,\bt) =  s^2 + t^2 - \al v(\bs,\bt),
\eeq
$\al = j/\jtil$, and $R_{2n}$ is a correction of order $2n$ is $s$ and $t$. For example, $R_4 = O(s^3t, s^2t^2, st^3)$.

We now expand the $\Phi^P$'s in \eno{pp_to_q} in Taylor series in $\bs$ and $\bt$, and evaluate the Gaussian integrals that
result. In this way, we find that correct to order $1/\jtil^2$,
\beq
\Phi^Q_{AB}
   = \exp\Bigl[ \frac{1}{2\jtil} \bigl( \grad_s^2 + \grad_t^2 +\grad_s\cdot\grad_t
                    + i \nhat_1\cdot(\grad_s \times \grad_t) \bigr)\Bigr] \Phi^P_A \Phi^P_B,
\eeq
and we must set $\bs = \bt = 0$ after differentiation. This expression may now be combined with the $\rmap$ operator to find the
Moyal product. While it is simple to do this in terms of the variables $\bs$ and $\bt$, it is ultimately more desirable to transform
to derivatives in terms of $\nhat$. Unfortunately, this transformation is extremely tedious to carry out beyond the lowest nontrivial
order. To this order, we can put $\bs \simeq \nhat_2 - \nhat_1$ and $\bt \simeq \nhat_3 - \nhat_1$, and
\beq
\Phi^W_{AB}(\nhat)
  = \Phi^W_A(\nhat) \Phi^W_B(\nhat)
    + \frac{i}{2\jtil} \nhat\cdot(\grad_{\nhat}\Phi^W_A \times \grad_{\nhat}\Phi^W_B)
    + O(\jtil^{-2}).
\eeq
The $O(1/\jtil)$ term may be written more physically in terms of the commutator $[A,B]$.
If we define $\jtil\nhat = \bj_c$ as the classical angular momentum, we have
\beq
\Phi^W_{[A,B]}(\bj_c) \apx  i\, \bj_c 
                  \!\cdot\! \Bigl( \frac{\ptl \Phi^W_A(\bj_c)}{\ptl\bj_c} \times \frac{\ptl \Phi^W_B(\bj_c)}{\ptl\bj_c}  \Bigr).
\eeq
But the right hand is precisely the Poisson bracket of two functions of $\bj_c$, given that the fundamental Poisson brackets are
$\{j_{c\al}, j_{c\be}\}_{{\rm PB}} = \eps_{\al\be\gam} j_{c\gam}$. Hence we have shown that to leading order as $j \to \infty$,
\beq
\Phi^W_{[A,B]} \apx i \{ \Phi^W_A, \Phi^W_B \}_{{\rm PB}}.
\eeq
Moreover, there is no correction to the anticommutator to this order, i.e.,
$
\Phi^W_{ \{A,B\} } \apx \Phi^W_A \Phi^W_B + O(\jtil^{-2})
$.

In summary, we have shown that the Weyl-Wigner-Moyal formulation of quantum mechanics in phase space may be carried out for
spin just as it can for position and momentum variables. We have derived exact asymptotic relations between the various representations
of an operator, and we have shown that for the commutator of two operators, the leading term is just the classical Poisson bracket.
These results should be of value in developing other semiclassical results for spin, for example for the spin-coherent-state
propagator.

\acknowledgments
This work was supported by the NSF via grant numbers PHY-0854896 and DGE-0801685 (NSF-IGERT program).
%

\vskip20pt
\noindent $^{\dagger}$Present address: Emma Willard School, 285 Pawling Ave., Troy, NY 12180

%
%

\begin{thebibliography}{99}
\bibitem{nonunique}
Note that \eno{def_psym} does not define $\Phi^P_A$ uniquely. The reason is that the matrix elements of
$\ket{\nhat}\bra{\nhat}$ in the $\ket{j,m}$ basis are linear combinations of $\ylm(\nhat)$ with $\ell \le 2j$. Because of the
orthogonality of the $\ylm$'s, we may add to $\Phi^P_A$ any linear combination of $\ylm$'s with $\ell > 2j$ without
affecting the integral over $\nhat$. In this paper we shall fix $\Phi^P_A$ by demanding that $\ylm$'s with
$\ell > 2j$ be absent.
\bibitem{easier}
The actual series may be found more easily by considering the action of $\rmap$ on $\plm(\nhat)$.
\end{thebibliography}
\end{document}